\documentclass[twocolumn,showpacs]{revtex4}      
\usepackage{wrapfig,epsfig,epsf}
\usepackage{dcolumn}
\usepackage{amssymb,amsmath}
\usepackage{mathtext}
\usepackage{rotate}
\usepackage{euscript}
\usepackage{bm}

\begin{document}

\input epsf
\renewcommand{\topfraction}{0.8}

\title {\Large\bf Some constraints on brane inflation models with power-law potentials }
 \author{\bf Sergey A. Pavluchenko}
\affiliation{ { Sternberg Astronomical Institute, Moscow State University, Moscow 119992, Russia \\
                and \\
                Department of Physics, University of Alberta, Edmonton, Alberta, Canada 
                }    }

{\begin{abstract}

We investigate inflation in Randall-Sundrum type II brane scenario with closed Friedman-Robertson-Walker (FRW) brane. 
We consider only power-law potentials of the scalar field and wide range of powers and parameters for them.
For our models we numerically calculate
the total number of e-folds, the value of potential at the end of inflation and amplitude and spectral index of scalar
perturbations at the epoch when the present Hubble scale leaves the horizon. All these values we calculate for different initial 
conditions and different values of
parameters. Then we compare our theoretical predictions with observation data and set constraints on the parameters of our model. 

\end{abstract}}
\pacs{98.80.Cq }

\maketitle

\section{Introduction}

Brane-world scenarios, after their discovering some years ago~(\cite{RS1,RS2}) quickly became
very popular. There are numerous issues in the field of branes and one of the most actual is brane-world inflation~\cite{chaot_br}. 
In this paper
we consider brane-world Randall-Sundrum type II (RSII)~\cite{RS2} scenario with power-law potential of the scalar field and study the 
possibility for successful
inflation in such model. Our current method is similar to method we used when studied inflation in closed 
Friedman-Robertson-Walker (FRW) models~\cite{my04}. Namely we calculate the total number of e-folds during inflation, the value
of potential at the end of inflation and the amplitude and spectral index of scalar perturbations at the epoch when present Hubble scale 
leaves the
horizon. Then we check if the total number of e-folds is larger than 72 and if it does we say the initial 
data for this model lead to inflation. This value must be larger than the number of e-folds between
the moment when present Hubble scale crosses the horizon during inflation and the moment of the end of inflation. This second value 
is model-dependent and it depends on 
some features of theory, such as the way of inflation ends etc.~\cite{Nhor}. It varies from theory to theory in a range from 55 to 75
approximately and using $N_{hor} = 62$ we use in some sense mean value for it. Our study shows that constraints one can set are slightly dependent on this
value if it changes in the range from 55 to 75. 

The second test is linked with the value of potential at the end of inflation. We calculate it and compare with observation 
data~\cite{V-constr}
$$
3 < \frac{V_{end}^{1/4}}{10^{15}\mbox{GeV}} < 29.    \eqno(1)
$$
\noindent For brane case the situation with the energy density at the end of inflation may be more complicated then in the FRW case 
(see~\cite{brane_end_infl} for details).

Also we can calculate index of the scalar perturbations spectrum at the epoch when present Hubble scale leaves the horizon during
inflation. Since there are observation constraints on scalar spectral index from WMAP~\cite{WMAP}, ACBAR~\cite{acbar}, CBI~\cite{cbi} 
and other CMB experimens~\cite{CMB-other} and large-scale structure~\cite{LSS}:
$$
n_S=0.99 \pm 0.04 \quad \mbox{(WMAP only)}  \eqno(2)
$$
 
$$
n_S = 0.97 \pm 0.03   \eqno(3)
$$

\noindent (WMAP+ACBAR+CBI+2dFGRS+$L_{\alpha}$-forest)
we can compare our predictions with these data and set some constraints on our model.

Finally we calculate the amplitude of scalar perturbations at the epoch when present Hubble scale leaves the horizon and compare it with 
COBE constraint~\cite{COBE} 
$$
A_S \sim 2\times 10^{-5}. \eqno(4)
$$

These three small tests help us in setting some constraints on the parameters of our model and on parameters of the potential. But in fact
we do not focus on one of parameters~-- {\it dark radiation}, because it actes only in the beginning of the inflation, so in this sence
it behaves itself something like curvature. The influence of the initial curvature on the total number of e-folds and background values for
FRW case was studied in~\cite{my04}
and we found that this influence is very weak. One can suspect this influence is weak in brane inflation as well.. 
But the dark radiation, unlike curvature, is one of the parameters which determine the initial value of the total energy
density, so it must act some differently.

The structure of the paper as follows. First, we write down the main equations we used in our work and describe our method. Then,
in Section III we describe the situation with quadratic potential, constraints on the parameters for this potential, in Section IV~--  the
same but for quartic potential, and linear potential is considered in 
Section V. Finally in Section VI we discuss our results.

\section{Main equations}

The main equations, which determine the evolution of our model, are~\cite{brane-eq1,brane-eq2,shtanov}

$$
\dot H - \frac{k}{a^2} = - \frac{4\pi}{m_4^2} (\rho + P) - \frac{2\pi}{3m_5^6} \rho (\rho + P) - \frac{2C}{a^4}, \eqno(5)
$$

$$
\ddot \varphi + 3 H \dot\varphi + \frac{dV(\varphi)}{dt} = 0, \eqno(6)
$$

\noindent and the first integral of our system is

$$
H^2 = \frac{8\pi}{3m_4^2} \rho + \frac{2\pi}{9m_5^6} \rho^2 + \frac{C}{a^4}, \eqno(7)
$$

\noindent where $m_4$ is 4D Planck mass and $m_5$ is 5D Planck mass. Below we use only $m_4$ so we need 
to write down the relation between these two masses:

$$
m_5^6 = \frac{\sigma m_4^2}{6}, \eqno(8)
$$

\noindent where $\sigma$ is brane tension. 

Since our aim is studing inflation and we are interesting in determination of the moment when inflation ends, let us rewrite
Eq.(5) in terms of $\ddot a$ and $\sigma$, and $m_4$ instead of $m_5$ using Eq.(8):

$$
\frac{\ddot a}{a} = \frac{k}{a^2} - \frac{1}{6m_4^2} (\rho + P) - \frac{1}{6\sigma m_4^2} \rho (2\rho + 3P) - \frac{C}{a^4}, \eqno(5*)
$$

\noindent or using normalization $m_4^2 = 8\pi$:

$$
\frac{\ddot a}{a} = \frac{k}{a^2} - \frac{1}{6} (\rho + P) - \frac{1}{6\sigma} \rho (2\rho + 3P) - \frac{C}{a^4}. \eqno(5**)
$$

Now we study brane inflation with power-law potential of the scalar field. Such potentials are well-known and well-studied in standard
FRW cosmology~\cite{PL-FRW,our01} and in brane models as well~\cite{chaot_br,PL-branes,0405490,0204115,0309608,0407543,0402126} and they lead to 'chaotic 
inflation'~\cite{chaotic}.
For power-law potentials we use following representation:

$$
V(\varphi) = \lambda \left( \frac{\varphi}{m_4} \right)^q. \eqno(9)
$$

As we noted above we use scalar spectral tilt as one of our tests. The most common view for it is~\cite{Huey-Lidsey}
$$
n_S -1 \approx -\frac{m_4^2 \lambda}{2\pi V(\varphi)} \left[ 3\frac{(V'(\varphi))^2}{V^2(\varphi)} - \frac{V''(\varphi)}{V(\varphi)} \right], 
$$

\noindent and using our normalization and Eq.(9) we can rewrite it as
$$
n_S = 1 - \frac{4\lambda}{V(\varphi)} \frac{n(2n+1)}{\varphi^2}.
$$

One more value which we use to set our constraints is the amplitude of scalar perturbations at the moment when present Hubble scale left the 
horizon during inflation. This value is given by~\cite{chaot_br,A_S,Huey-Lidsey,0402126}
$$
A_S^2 \simeq \left( \frac{512\pi}{75 m_{4}^6}  \right) \frac{V^3(\varphi)}{(V'(\varphi))^2} \left[ \frac{2\lambda + V(\varphi)}{2\lambda} 
\right]^3.
$$

\noindent We can rewrite it using Eq.(9) and our normalization $m_4^2 = 8\pi$:
$$
A_S^2 \simeq \frac{1}{75\pi^2} \frac{\lambda \varphi^{q+2}}{m_{4}^q q^2} \left[ \frac{2\lambda + V(\varphi)}{2\lambda} 
\right]^3.
$$

And finally about our method. Like in our previous papers about inflation~\cite{my04,my03,our01}, we start from Planck boundary and than integrate
equations (5**),(6) through inflation. Also we check constraint Eq.(7) to do not diverge. Like in~\cite{our01} we start not
from 4D but from 5D Planck boundary. So from left-hand side of Eq.(7) one can see:

$$
H^2 + \frac{1}{a^2} = m_5^2,
$$

\noindent and one can parametrize initial $a$ and $\dot a$ by next way:

$$
H_0 \in [0;m_5); \; a_0 = \frac{1}{\sqrt{(m_5^2 - H_0^2)}}; \; \dot a_0 = \frac{H_0}{\sqrt{(m_5^2 - H_0^2)}}.
$$

The value $(H_0/m_5)$ completely parametrizes initial $a$ and $\dot a$. From right-hand side of Eq.(7) one can see that there are three 
extra parameters: $\sigma$~--
brane tension, $C$~-- dark radiation and last parameter is $\alpha$~-- the ratio of kinetic energy of the scalar field to the total energy 
density of the scalar field. Last value allows us to calculate initial values for $\varphi$ and $\dot\varphi$. So finally we have five
parameters: $(H_0/m_5)$, which describes initial curvature; $\alpha$, which describes a contribution of the kinetic energy to the total
energy density of the scalar field. These two parameters describe initial conditions, last three are parameters of the model: brane tension 
$\sigma$, parameter from the potential $\lambda$ and dark radiation $C$. Requirement that model should to be consistent with Newton's Law at
small distances sets constraint on $\sigma$~\cite{chaot_br,sigma-const,Huey-Lidsey}:

$$
m_5 > 10^5 \mbox{TeV} \quad \mbox{or} \quad \sigma > 10^8 \mbox{GeV}^4. \eqno(10)
$$

Also from right-hand side of Eq.(7) one can set constraint on initial value of $C$:

$$
C \leqslant m_5^2 a_0^4  \eqno(11)
$$

\noindent (in order to energy density of dark radiation do not exceed 5D Planck boundary). And let us note~-- due to pure geometrical
nature of this dark energy term its energy density need not to be positive. And as we can see from Eq.(11) value of $C$ is bounded upper
but unbounded below.

\begin{figure*}
\epsfxsize=18cm
\centerline{{\epsfbox{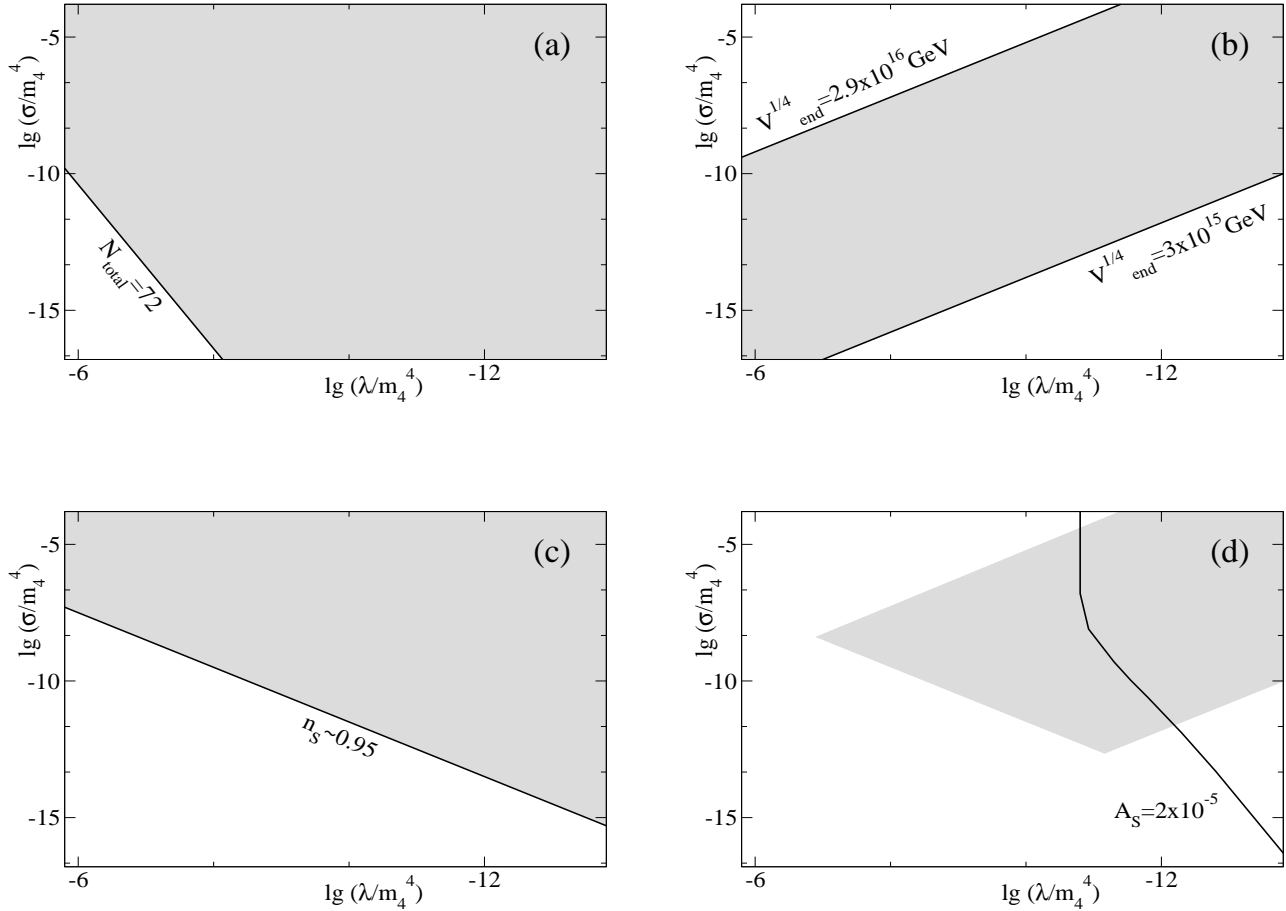}}}
\caption{Constraints on both $\lg(m)$ and $\lg(\sigma)$ for $q=2$ case: from $N_{total} > 72$ only in (a); from Eq.(1) only in (b);
from constraints on $n_S$ in (c) and all previous combined plus constraint from $A_S$ in (d) (see text for details).}
\end{figure*}

\section{Quadratic potential}

In this section we present our results for case of quadratic potential. They are presented in Fig. 1. 

In Fig. 1(a) we present only a
region with $N_{total} > 72$ (grey area). In fact most of trajectories (as a {\it trajectory} we mean an evolution curve for a model with 
particulair initial conditions in some coordinates, say $(a(t),t)$ {\it etc.}) with initially large enough $\rho^2$-term show oscillationary
behavior (this is due to so-called duality between low-energy Friedmann regime and high-energy brane $\rho^2$-regime~\cite{0405490,PL-branes}), so
as the total number of e-folds we mean the maximal number of e-folds before first $\ddot a > 0$ violation. 

In Fig. 1(b) we present only a region with parameters which lead to values of the potential at the end of inflation which obey Eq.(1) (grey
region).
In Fig. 1(c) models with $m$ and $\sigma$ leading to $n_S > 0.95$ are shown. One can see from Eq.(9) that always $n_S < 1$. In fact one
can choose as a boundary value for $n_S$ not 0.95 but say 0.94 (see Eqs. (2) and (3)), but these two values are practically indistinguishible and they both
lead to practically similair constraints on $m$ and $\sigma$, so we use single value $n_S = 0.95$.
In Fig. 1(d) we represent all constraints from (a) to (c) together with last constraint linked with Eq.(2). 
Parameters from grey area in $(\lg (m), \lg(\sigma))$-parameter space in Fig. 1(c) obey requirements from all (a), (b) and (c)~-- their total
number of e-folds is larger than 72, the value of the potential and the end of inflation lies in a range of Eq.(1) and $n_S < 0.95$. 
Bold black line
in Fig. 1(d) corresponds to parameters with amplitude of the spectrum of scalar perturbations at the epoch when the present Hubble scale 
left the horizon obeys Eq.(2). Let us remind the reader we choose this value as 62, i.e. the present Hubble scale left the horizon 62 e-folds
before the end of inflation. In fact one may choose another value, say, in~\cite{my04} we used two different values~-- 62 and 55 and we
found that there is no significant difference between these two cases, see~\cite{my04} for details.

\begin{figure*}
\epsfxsize=18cm
\centerline{{\epsfbox{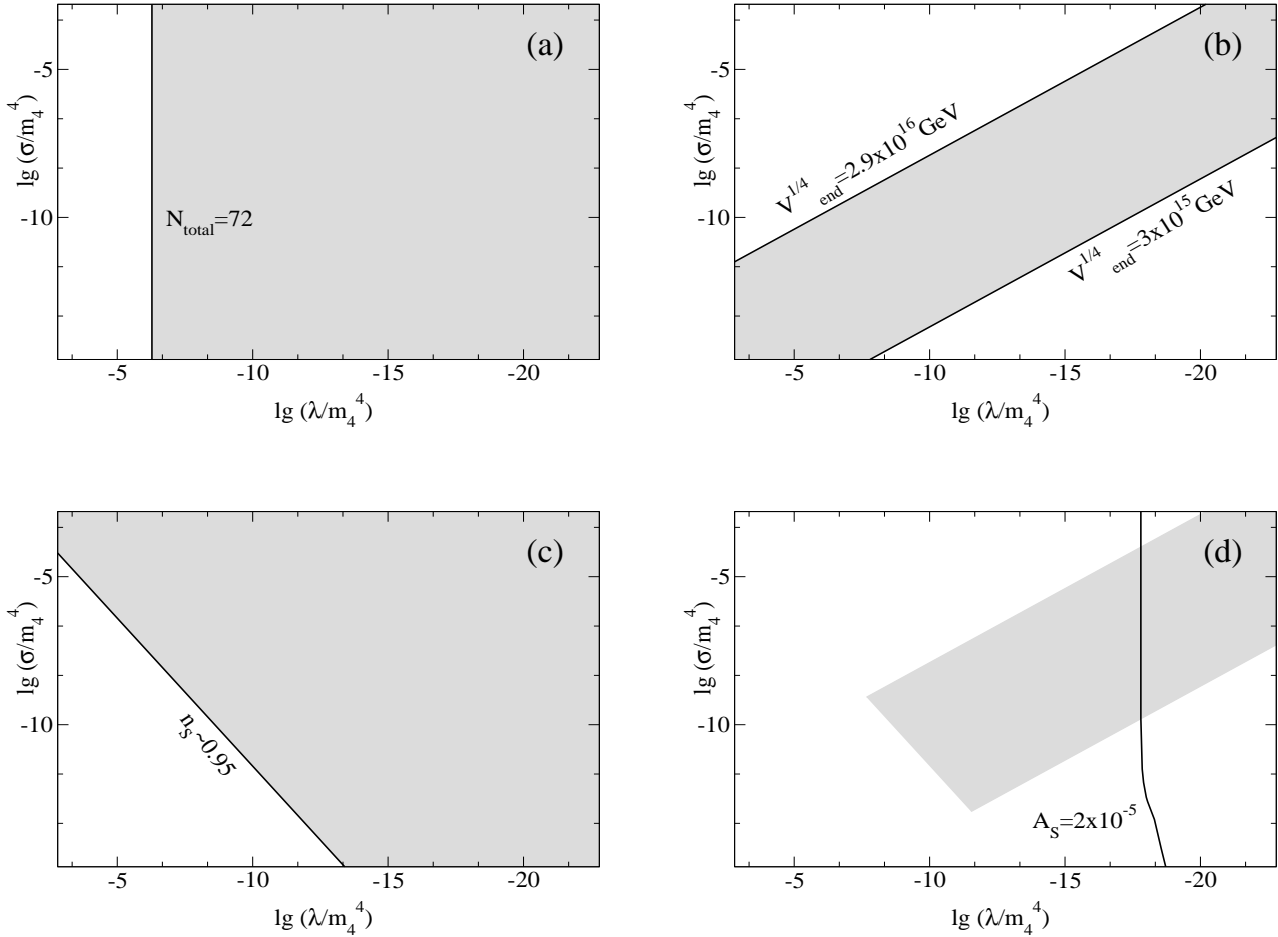}}}
\caption{Constraints on both $\lg(m)$ and $\lg(\sigma)$ for $q=4$ case: from $N_{total} > 72$ only in (a); from Eq.(1) only in (b);
from constraints on $n_S$ in (c) and all previous combined plus constraint from $A_S$ in (d) (see text for details).}
\end{figure*}

Finally, from Fig. 1(c) one can set constraints on both $m$ and $\sigma$ or on one of them with fixed another one. Values for $\sigma$ obey 
Eq.(10). From intersection of curve $A_S=2\times 10^{-5}$ with grey area one can set constraint on $\lambda$: $\lambda \in (2.5\times 10^{-13}; 
 1.6 \times 10^{-11}) m_4^4$. 
And constraint on $\sigma$ is: $\sigma \in (2\times 10^{-12}; 4\times 10^{-5}) m_4^4$.

\section{Quartic potential}

This section is devoted to the quartic potential and our results are presented in Fig. 2. One can see that some global changes occur from the 
$q=2$ case. Say, the requirement $N_{total} > 72$ leads to simple requirement $\lambda > 3.3\times 10^{-4}$ in $q=4$ case (see Fig.2(a)), and in 
$q=2$ case it leads to some more complicated constraint dependent on both $\lambda$ and $\sigma$ (see Fig.1(a)). Area, which one can get from 
Eq.(2),
is shown in Fig. 2(b). It differs from Fig. 1(b) in another scope in $(\lg \lambda; \lg \sigma)$ coordinates. All these changes are due
to differ in powers of the power-law potential in cases $q=2$ (Fig. 1) and $q=4$ (Fig. 2).
In Fig. 2(c) we represent 
constraints on $\lambda$ and $\sigma$ from Eqs.(2) and (3). It remains practically unchanged from $q=2$ case. 
And finally in Fig. 2(d) we
summarized all constraints from Figs. 2(a) to 2(c) and add a curve which corresponds to Eq.(4). So from Fig. 2(d) we can set following 
constraints on the parameters of our model: $\lambda \sim 1.6\times 10^{-18} m_4^4$, $\sigma \in ( 1.6\times 10^{-10};  10^{-5}) m_4^4$.

\begin{figure*}
\epsfxsize=18cm
\centerline{{\epsfbox{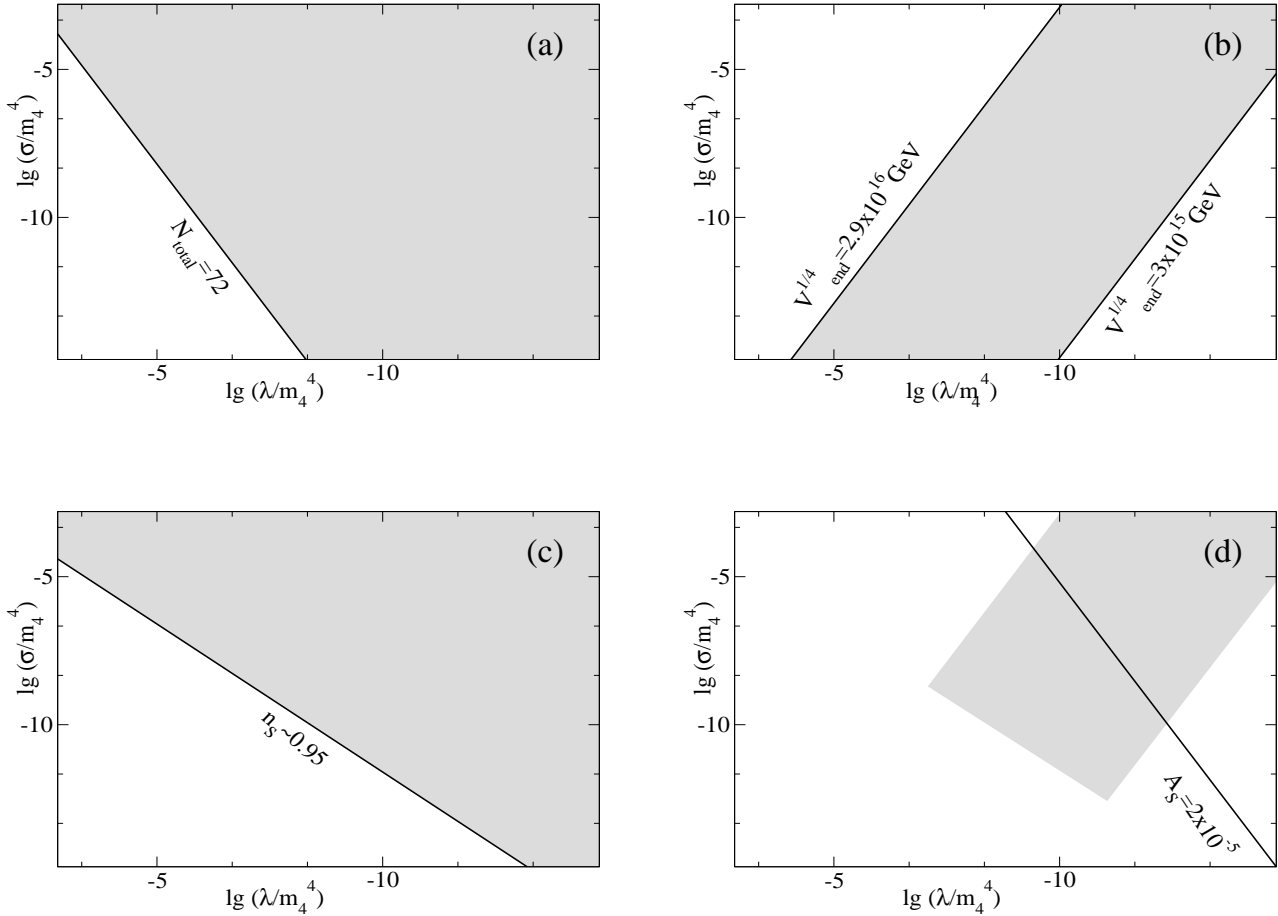}}}
\caption{Constraints on both $\lg(m)$ and $\lg(\sigma)$ for $q=1$ case: from $N_{total} > 72$ only in (a); from Eq.(1) only in (b);
from constraints on $n_S$ in (c) and all previous combined plus constraint from $A_S$ in (d) (see text for details).}
\end{figure*}

\section{Linear potential}

Finally we consider linear potential and our results are presented in Fig. 3. Analogically with two previous cases in Fig. 3(a) we represent
on the parameter space values of parameters which lead to inflation with total number of e-folds more than 72, in Fig. 3(b) those who
obey Eq.(1), in Fig. 3(c)~-- those who obey $n_S >~ 0.95$ and in Fig. 3(d) we summarized three previous constraints and add a curve which
corresponds to Eq.(4). So from Fig. 3(d) one can set constraints on $\lambda$ and $\sigma$: $\lambda \in ( 5\times 10^{-13}; 5\times 10^{-10} 
) m_4^4$; $\sigma \in (5\times 10^{-11}; 5\times 10^{-13})m_4^4$.

\section{Summary and discussion}

We have investigated three cases for brane inflation with power-law potential of the scalar field. Namely we considered linear, quadratic and
quartic potentials and found constraints on the parameters for these three particulair cases. Constraints on the parameters for these three
models are summarized in Table 1. And now let us generalize our results.

\begin{table}
\caption{\label{tab:table1}In this table we summarized our constraints on brane inflation models with power-law potentials for three
different powers.}
\begin{ruledtabular}
\begin{tabular}{lccc}
 & $q=1$ & $q=2$ & q=4 \\
\hline
$\lg \sigma$ & $-4.3 ... -10.3$ & $-4.4 ... -11.7$ & $-4.0 ... -9.8$\\
\hline
$\lg \lambda$ & $-9.3 ... -12.3$ & $-10.8 ... -12.6$ & $\sim -17.8$\\
\end{tabular}
\end{ruledtabular}
\end{table}

From comparing Figs. 1, 2 and 3 we can learn that for all three cases we have similair slope of the curve $n_S$. Slopes of
lines $N_{total} = 72$ and lines correspondent to Eq.(1) are different due to different powers. But the 'best' behavior (from the model's
testability point of view) shows constraint from Eq.(4). Namely increasing of the power leads to decreasing of the maximal value of $\lambda$ 
which
corresponds to increasing power. This means for $q>4$ we always have an area from intersection of $N_{total}$, $n_S$ and $V_{end}$ 
constraints and this area always is intersected by $A_S$ line. The only constraint on maximal value of the power one can set from the fact that
the maximal value of $\lambda$ need obey Eq.(10).

Let us note that one can also set some constraints on this model but from other point of view. Say, in~\cite{0307017,0312162,0407543} some 
constraints made from some other background. But apart from our results they found that even quartic potential is under strong pressure
from observation. Let us also note that in Friedmann case our~\cite{my04} results and results from~\cite{0306305} are similair~-- both of them
lead to $q~<4$ constraint. 

And last our example is linear potential. This is very uncommon potential for inflation, but for brane inflation case it works (see 
also~\cite{0312162}). But for case $q<1$ there simple may occur that there will be no intersection between final area of $N_{total}$, 
$n_S$ and $V_{end}$ with $A_S$ curve: $A_S$ curve seems to 'move' to high values of $\lambda$ with decreasing of power and at some low
enough power it can lie at $\lambda$'s that higher than any $\lambda$ of the area of final constraints from $N_{total}$, 
$n_S$ and $V_{end}$.

\section{Acknowledgements}

We want to thank N. Savchenko for useful discussions. 
This work is supperted by the Russian Ministry of Industry, Science and Technology through the Leading Scientific School Grant \#2338.2003.2.

\end{document}